\documentclass[aps,prb,preprint,superscriptaddress]{revtex4-1}

\usepackage{graphicx}
\usepackage{dcolumn}
\usepackage{tabularx}
\usepackage{gensymb}
\usepackage[colorinlistoftodos]{todonotes}

\usepackage{ulem}

\begin{document}

\title{Electronic Band Dispersion of Graphene Nanoribbons via Fourier-Transformed Scanning Tunneling Spectroscopy}

\author{Hajo S\"{o}de}
\author{Leopold Talirz}
\author{Oliver Gr\"{o}ning}
\affiliation {Empa, Swiss Federal Laboratories for Materials Science and Technology, nanotech$@$surfaces Laboratory, \"Uberlandstrasse 129, CH-8600 D\"{u}bendorf, Switzerland}
\author{Carlo Antonio Pignedoli}
\affiliation {Empa, Swiss Federal Laboratories for Materials Science and Technology, nanotech$@$surfaces Laboratory, \"Uberlandstrasse 129, CH-8600 D\"{u}bendorf, Switzerland}
\affiliation {Empa, Swiss Federal Laboratories for Materials Science and Technology, NCCR MARVEL, \"Uberlandstrasse 129, CH-8600 D\"{u}bendorf, Switzerland}
\author{Reinhard Berger}
\author{Xinliang Feng}
\author{Klaus M\"{u}llen}
\affiliation {Max Planck Institute for Polymer Research, 55124 Mainz, Germany}
\author{Roman Fasel}
\affiliation {Empa, Swiss Federal Laboratories for Materials Science and Technology, nanotech$@$surfaces Laboratory, \"Uberlandstrasse 129, CH-8600 D\"{u}bendorf, Switzerland}
\affiliation {Department of Chemistry and Biochemistry, University of Bern, Freiestrasse 3, CH-3012 Bern, Switzerland}
\author{Pascal Ruffieux}
\email[]{pascal.ruffieux$@$empa.ch}
\affiliation {Empa, Swiss Federal Laboratories for Materials Science and Technology, nanotech$@$surfaces Laboratory, \"Uberlandstrasse 129, CH-8600 D\"{u}bendorf, Switzerland}

\date{\today}

\begin{abstract}

The electronic structure of atomically precise armchair graphene nanoribbons of width $N=7$ (7-AGNRs) are investigated by scanning tunneling spectroscopy (STS) on Au(111).
We record the standing waves in the local density of states of finite ribbons as a function of sample bias and extract the dispersion relation of frontier electronic states by Fourier transformation.
The wave-vector-dependent contributions from these states agree with density functional theory calculations, thus enabling the unambiguous assignment of the states to the valence band, the conduction band and the next empty band with effective masses of $0.41\pm 0.08\,m_e$, $0.40\pm 0.18\,m_e$ and $0.20\pm 0.03\,m_e$, respectively.
By comparing the extracted dispersion relation for the conduction band to corresponding height-dependent tunneling spectra, we find that the conduction band edge can be resolved only at small tip-sample separations and has not been observed before. As a result, we report a band gap of $2.37\pm 0.06$~eV for 7-AGNRs adsorbed on Au(111). 

\end{abstract}


\maketitle

\section{Introduction}
It was predicted as early as 1996 that tailoring graphene into nanometre-wide ribbons, termed graphene nanoribbons (GNRs), would give rise to electronic properties that differ strongly from those of the semimetallic parent material \cite{Nakada1996}.
These properties include sizable electronic band gaps due to quantum confinement and edge effects \cite{Yang2007}, as well as the spatial separation of spin channels due to spin-polarized  edge states in zigzag GNRs \cite{Wakabayashi2010,Wimmer2008}. 
However, the electronic structure of GNRs sensitively depends on their specific atomic configuration and the fabrication of GNRs with atomic precision was made possible only recently through advances in bottom-up approaches \cite{Cai2010,Chen2013,Narita2014}. In the route chosen in this work, specifically designed precursor monomers are deposited onto a metal surface, followed by their surface-assisted colligation and subsequent cyclodehydrogenation \cite{Cai2010}.

\begin{figure}[h]
\includegraphics[width=0.75\textwidth]{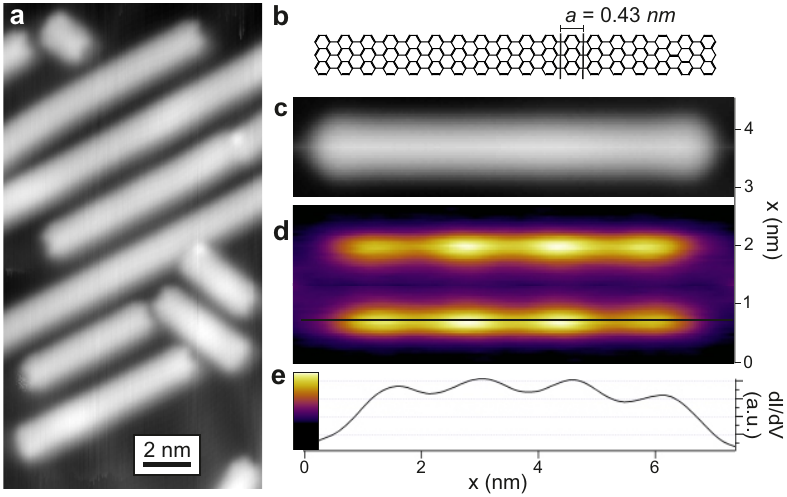}
\caption{\label{fig:1} (a) STM image of short 7-AGNRs on Au(111) ($V=-0.5$~V, $I=0.05$~nA). (b) Schematic model of 7-AGNR investigated in (c-e), with a length of 18 unit cells ($18a\approx 7.7$~nm). (c) Constant current STM topography image ($V=-1.3$~V, $I=0.5$~nA), and (d) conductance map recorded simultaneously. (e) Differential conductance line profile along the path indicated and color scale used in (d).}
\end{figure}

First electronic structure studies have concentrated on the armchair GNR (AGNR) with a width of $N=7$ carbon dimer lines (7-AGNR), which is shown in Figure~\ref{fig:1}. Its electronic band gap $E_g$ as well as the effective masses $m^*_{VB}$ of the  valence band and $m^*_{CB}$ of the conduction band have been investigated using different methods, arriving at conflicting results.
A band gap $E_g$ of $2.3$~eV \cite{Ruffieux2012}, resp. $2.5$~eV \cite{Chen2013}, was determined by scanning tunneling spectroscopy for the 7-AGNR adsorbed on Au(111).
Angle-resolved photomission spectroscopy (ARPES) experiments on 7-AGNRs aligned on vicinal Au surfaces provided values for the effective mass $m^*_{VB}=0.21\,m_e$ \cite{Ruffieux2012} and $1.07\,m_e$ \cite{Linden2012}, where $m_e$ is the free electron mass. We note that GNRs were grown ex-situ in both works, but other than that both papers were not able to explain the massive difference between the two values and the ones expected from DFT simulations. By combining ARPES with inverse PES, Linden et al. arrived at a band gap of $2.8$~eV  \cite{Linden2012}.
Finally, from angle-resolved two-photon photoemission spectroscopy (2PPE) effective masses of $m^*_{VB}=1.37\,m_e$ and $m^*_{CB}=1.35\,m_e$ were extracted \cite{Bronner2014}, though the 7-AGNRs under study were oriented randomly.

In view of the discrepancy between these results, determining a consistent set of parameters by the same in-situ experimental technique is highly desirable. 
Here, we report band masses and band onsets for the VB, CB and CB+1 for the 7-AGNR on Au(111), determined by Fourier-transformed scanning tunneling spectroscopy (FT-STS). 
We emphasize that the analyzed 7-AGNRs are free of defects and of sufficient length. The precise knowledge of their atomic structure enables direct comparison with ab initio electronic structure calculations.

\section{Methods}

\subsection{Fourier-transformed Scanning Tunneling Spectroscopy of GNRs}

The physical quantity measured in STS is the derivative d$I$/d$V$ of the tunneling current $I$ with respect to the sample bias $V$.
In the Tersoff-Hamann approximation \cite{Tersoff1985}, d$I$/d$V(V,\vec{r})$ at sample bias $V$ and tip position $\vec{r}$ is proportional to the local density of states (LDOS) $\rho(E,\vec{r})$ at energy $E=E_F+|e|V$, where $E_F$ denotes the Fermi energy. STS maps like the one shown in Fig.~\ref{fig:1} (d) can thus be interpreted as maps of the local density of states.

In a perfect crystal, the electronic Bloch wave functions in neighboring unit cells differ only by a phase factor, resulting in a local density of states that shares the periodicity of the underlying lattice. Defects, however, give rise to scattering of Bloch waves, leading to interference patterns with characteristic wave vectors.
By mapping out these standing waves in the LDOS as a function of position and sample bias, STS can be used to reconstruct the energy-momentum relation for both occupied and empty electronic states\cite{Burgi2000,Bergvall2013}.
Systems studied so far by this approach include defects in graphene \cite{Andrei2012}, carbon nanotubes \cite{Buchs2009} and high-$T_c$ superconductors \cite{Balatsky2006}, but also the ends of carbon nanotubes \cite{Bercioux2011} as well as short polyphenylene chains \cite{Wang2011,Wang2012a}.

\begin{figure*}
\includegraphics[width=0.98\textwidth]{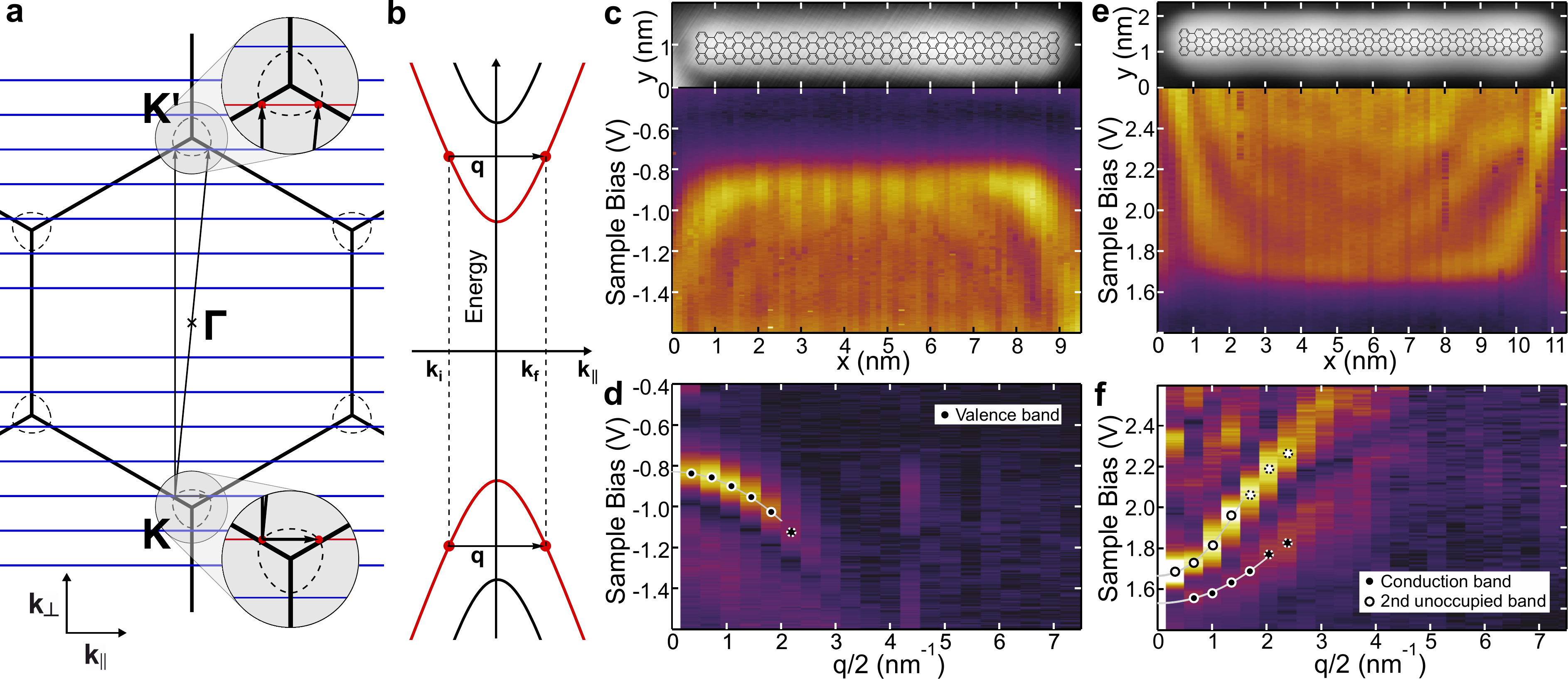}
\caption{\label{fig:2} 
(a) Reciprocal lattice of graphene with discretized $k_\perp$ spectrum of the 7-AGNR \cite{Wakabayashi2010}, showing constant-energy contours (dashed) and possible scattering processes in 2D momentum space.
(b) Scattering processes from (a) in $E$ versus $k_\parallel$ space.
(c) STM image and equidistant d$I$/d$V(V)$ spectra of occupied states, recorded along the edge of a 7-AGNR of length $20a$  (set point $V=-1.6$~V, $I=0.3$~nA, spacing $\delta x=0.12$~nm).
(d) Line-by-line Fourier transform of (c) for $0\leq \frac{q}{2} \leq \frac{\pi}{a}$, including parabolic fit near $q/2=0$.
(e) Analogous STM and d$I$/d$V(V)$ data for the unoccupied states, recorded along the edge of a 7-AGNR of length $24a$ (set point $V=2.7$~V, $I=0.6$~nA, spacing $\delta x=0.11$~nm). 
(f) Line-by-line Fourier transform of (e).}
\end{figure*}

In the case of armchair GNRs, elastic scattering occurs both at the long armchair edges and at the short zigzag edges, giving rise to selection rules and standing wave patterns \cite{Bergvall2013}.
To a first approximation, the frontier electronic structure of armchair GNRs can be described by a single-orbital nearest-neighbor tight binding model, considering only the carbon $\pi$-electrons \cite{Wakabayashi2010}.
Scattering at the armchair edges then restricts the wave vector component perpendicular to the GNR axis to discrete values $k_\perp = \pm\frac{2\pi}{a/\sqrt{3}}\frac{r}{N+1},\ r\in\{1\ldots N\}$, where $N$ denotes the width of the AGNR.
As illustrated for $N=7$ in Fig.~\ref{fig:2} (a), the one-dimensional band structure of armchair GNRs can be obtained simply by cutting the two-dimensional band structure of graphene at the indicated values of $k_\perp$. 
Due to the small width of the 7-AGNR of $\approx 1$~nm, the scattering wave vectors are large, making it difficult to resolve the corresponding standing waves in real space.

Here, we focus instead on the scattering at the short zigzag edges, i.e. at the termini of finite 7-AGNRs. As illustrated in Figure~\ref{fig:2} (a-b), a left-moving initial Bloch state with $k_\parallel=k_i$ is scattered into a right-moving final state with $k_\parallel=k_f$, where $|k_i|=|k_f|=:k$ due to symmetry. The superposition of these two states gives rise to a standing wave in the local density of states with non-vanishing  Fourier components at the scattering vectors $\pm q$, where $q=|k_f-k_i|=2k$.

By scanning along the axis of the 7-AGNR as indicated in Fig.~\ref{fig:1} (d), we record the standing wave patterns in the LDOS $\rho(E,x)$. After acquisition of the line scans, we perform a discrete Fourier transform $\rho(E,x) \rightarrow \hat{\rho}(E,q)$ defined by

\[\hat{\rho}\left(|e|V, \frac{2\pi n}{M\delta x}\right)
  \propto \sum\limits_{m=0}^{M-1} \frac{dI}{dV}(V,m \delta x) \exp\left(i\frac{2\pi n}{M} m\right), \]
with the number of spectra $M$.By following the intensity maxima in $|\hat{\rho}(E,q)|$ as a function of $k=q/2$, the band dispersion of occupied and unoccupied states can be extracted.
We note that this approach involves the selection of a real-space window of length $L=M\delta x$, defining the spacing $\delta q = \frac{2\pi}{L}$ of the reciprocal grid. The choice of the grid affects the Fourier transform to a certain extent and the resulting uncertainty in the band masses is incorporated into the error bars.

\subsection{Experimental setup}

Measurements were performed at $5$~K in a commercial (Omicron) ultrahigh vacuum low-temperature STM setup. 
10,10'-dibromo-9,9'-bianthryl (DBBA) molecules were deposited by thermal sublimation onto Au(111) surfaces.
The 7-AGNRs were grown following the procedure reported by Cai et al.\ \cite{Cai2010}. The temperatures for polymerization and cyclodehydrogenation were lowered to 160\degree C and 320\degree C, respectively, yielding shorter GNRs. STS experiments were performed on GNRs with lengths in the range of $8-16~\textrm{nm}$.
The GNRs showed either H$_1$ or H$_2$ terminations at the central carbon atom of the short zigzag edge \cite{Talirz2013}. We verified, however, that the band gap of 7-AGNRs with lengths above $8$~nm approaches the one of long 7-AGNRs to within $0.05$~eV, independent of their termination.

Standing wave patterns along the 7-AGNRs were recorded by taking differential conductance spectra at equidistant points on a topography line-scan along the GNR (d$I$/d$V(V,x)$).
We note that the choice of the grid-spacing $\delta x$ is fine enough not to limit the resolution in reciprocal space, i.e. $k_{max} = \frac{\pi}{2 \delta x}\gg \frac{\pi}{a}$.
Since the LDOS is highest near the edge of the GNRs (as rationalized in the discussion), line-scans were performed along the edges as illustrated in Figure~\ref{fig:1} (d).
The STM bias voltage was modulated at $860$~Hz with $20$~mV and the differential conductance was measured by lock-in technique.
The set points for current and bias voltage required for the conductance spectra prevented us from acquiring high-resolution images simultaneously. However, the quality of the measured GNRs was checked beforehand with different current and bias voltage settings that reduced the tip-sample distance and improved resolution.

\subsection{Computational setup}

Density functional theory (DFT) calculations of 7-AGNRs in vacuum were performed using the PBE generalized gradient approximation to the exchange-correlation functional \cite{Perdew1996}. Structures were relaxed  until the forces acting on the atoms were below $3$~meV/\AA .
For finite ribbons we used the CP2K code \cite{cp2k}, which expands the wave functions on an atom-centered Gaussian-type basis set. 
After extrapolating the electronic states into the vacuum region \cite{Tersoff1989}, STS simulations were performed in the Tersoff-Hamann approximation \cite{Tersoff1985} on a plane parallel to the planar GNR, using a Lorentzian broadening of full-width $150$~meV  at half-maximum.
Band structure and effective masses of the infinite 7-AGNR were calculated with the Quantum ESPRESSO package \cite{Giannozzi2009} using norm-conserving pseudopotentials, a cutoff of $150$~Ry for the plane wave basis of the wave functions and a grid of $128$ k-points in the first Brillouin zone.

\section{Results and Discussion}

\newcolumntype{C}{>{\centering\arraybackslash}X}

\subsection{Band dispersion}
Fig.~\ref{fig:2} (c) and (e) show the LDOS $\rho(E,x)$ of occupied and empty states measured on defect-free 7-AGNRs.
In Fig.~\ref{fig:2} (d) and (f) their Fourier transform $|\hat{\rho}(E,q)|$ (FT-LDOS) is shown for $0<q/2<\frac{\pi}{a}$ corresponding to the first Brillouin zone of the 7-AGNR. In both measurements the Fourier transform reveals dispersing bands. 

In order to extract the effective masses $m^*$ and band onsets $E(k=0)$, we perform a two-parameter parabolic fit
\[ E(k) = E(k=0) + \frac{\hbar^2}{2 m^*}k^2 \]
to the data points shown in Fig.~\ref{fig:2} (d) and (f) near $k=0$.

For the valence band (VB), we obtain an effective mass of $m_{VB}^*=0.41 \pm 0.08\,m_e$ and a band onset at $E_{VB}(k=0)=-0.84 \pm 0.05$~eV with respect to the Fermi energy.
Corresponding results for the two lowest unoccupied bands have been extracted from Fig.~\ref{fig:2} (f), namely the effective masses $m_{CB}^*=0.40 \pm 0.18\,m_e$ and $m_{CB+1}^*=0.21 \pm 0.02\,m_e$, and the band onsets $E_{CB}(k=0)=1.52 \pm 0.02$~eV and $E_{CB+1}(k=0)=1.67 \pm 0.02$~eV.
We note that the lack of intensity for $k > 4$~nm$^{-1}$ indicates a resolution limit due to finite tip-sample separation and tip size, and does not imply the absence of scattering with wave vectors $k > 4$~nm$^{-1}$.

\begin{table}[h]
\bgroup
\def\arraystretch{1.8}
\begin{tabularx}{0.8\textwidth}{r|C|C|C}
 & VB & CB & CB+1 \\ \hline
$E(k=0)$ [eV]  & $-0.84\pm 0.05$ & $1.52\pm 0.04$  & $1.67\pm 0.03$ \\ \hline
$m^*\ [m_e]$  & $0.41\pm 0.08$ ($0.33$) & $0.40\pm 0.18$ ($0.41$) & $0.20\pm 0.03$ ($0.14$)
\end{tabularx}
\caption{\label{tab:1}
Band extrema and effective masses for the different bands averaged over several measurements on 7-AGNRs of length between $9$~nm and $16$~nm. DFT results in brackets.
}
\egroup
\end{table}

Table~\ref{tab:1} summarizes the band onsets and effective masses averaged over several measurements on different 7-AGNRs (see Appendix A). 
The band masses are compared with band masses from DFT calculations of the freestanding 7-AGNR (in brackets). 
We note that this comparison is justified by the fact that 7-AGNRs interact only weakly with the Au(111). Treating the substrate explicitly within DFT has been demonstrated to have negligible effects on the band dispersion\cite{Ruffieux2012}. And while many-body effects lead to a substantial opening of the DFT band gap\cite{Ruffieux2012}, the effect on band dispersion is expected to be only of the order of 10-20\%\cite{Gruneis2008}.

Given this, we find the measured effective masses to be in good qualitative agreement with the corresponding DFT values.
On the other hand, the valence band masses of $m^*_{VB}=0.21\,m_e$ \cite{Ruffieux2012} and $1.07\,m_e$ \cite{Linden2012} determined by ARPES and $m^*_{VB}=1.37\,m_e$ \cite{Bronner2014} determined by 2PPE  all deviate significantly from the predicted values.
The origin of this discrepancy is not understood at present and will be investigated in future work.

\subsection{Effect of tip-sample distance}

\begin{figure}[h]
\includegraphics[width=0.65\textwidth]{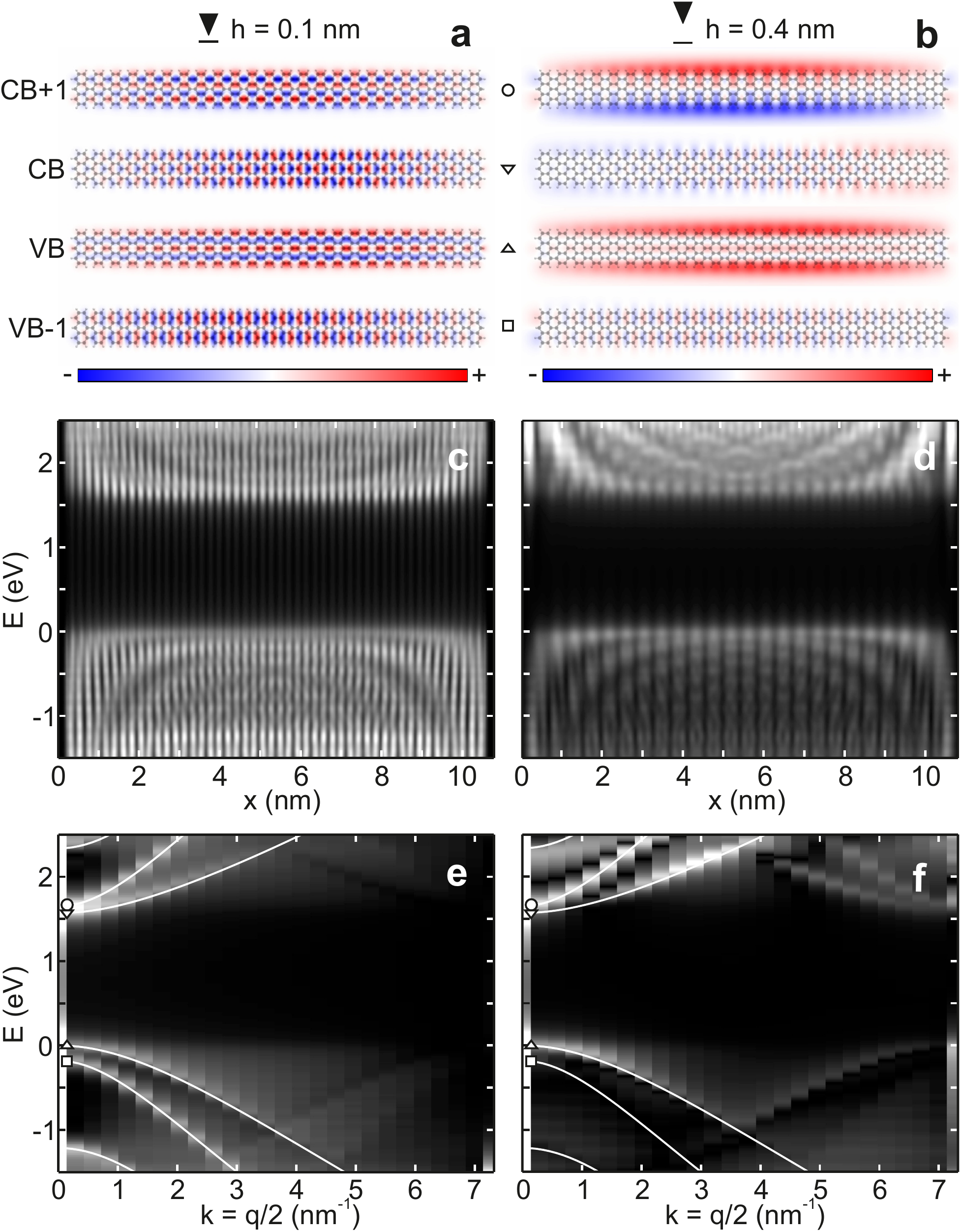}
\caption{\label{fig:3} Effect of tip-sample distance on probed LDOS. (a, b) Kohn-Sham orbitals at the band onsets for a 7-AGNR of length $24a$ evaluated $0.1$~nm (a) and $0.4$~nm (b) above the GNR. (c, d) LDOS along the 7-AGNR, integrated across the ribbon. $E_VB(k=0)$ is set to zero. (e, f) FT-LDOS for $0\leq k \leq \frac{\pi}{a}$ with bands of infinite 7-AGNR superposed as white lines.}
\end{figure}

In the FT-LDOS of the empty states shown in Fig.~\ref{fig:2} (f), the CB appears much fainter than the CB+1, and its signal further looses intensity towards $q/2=0$. In the FT-LDOS of the occupied states (Fig.~\ref{fig:2} (d)) only one valence band is detected, although DFT predicts a second valence band close by.
In the following, we rationalize these observations by studying the shape of the corresponding electronic orbitals in DFT.

Figure~\ref{fig:3} shows the Kohn-Sham orbitals of a finite 7-AGNR at the respective band onsets. The orbitals have been evaluated once on a plane $0.1$~nm above the GNR, corresponding to a short tip-sample distance (a), and once at a more realistic distance of $0.4$~nm (b). 
While the orbitals arising from the CB and the VB-1 oscillate strongly along and perpendicular to the ribbon axis, the orbitals arising from the VB and CB+1 change sign only in the direction perpendicular to the ribbon axis.
As a consequence, the LDOS arising from the CB and VB-1 decay faster as a function of tip-sample distance than their VB and CB+1 counterparts.
We now follow the same procedure as in experiment, by first calculating the LDOS along the 7-AGNR (c-d) and then taking its Fourier transform (e-f).
In Fig.~\ref{fig:3} (e), corresponding to $0.1$~nm tip-sample distance, all band onsets can be clearly determined from the FT-LDOS. In Fig.~\ref{fig:3} (f) at $0.4$~nm distance, however, the VB-1 and VB-2 are missing completely and the intensity due to the CB is strongly reduced at low $k$ values. 
By comparison with Fig.~\ref{fig:2} (d,f) we can therefore confidently label the experimentally resolved bands as the VB, CB and CB+1.

\begin{figure}[h]
\includegraphics[width=0.6\textwidth]{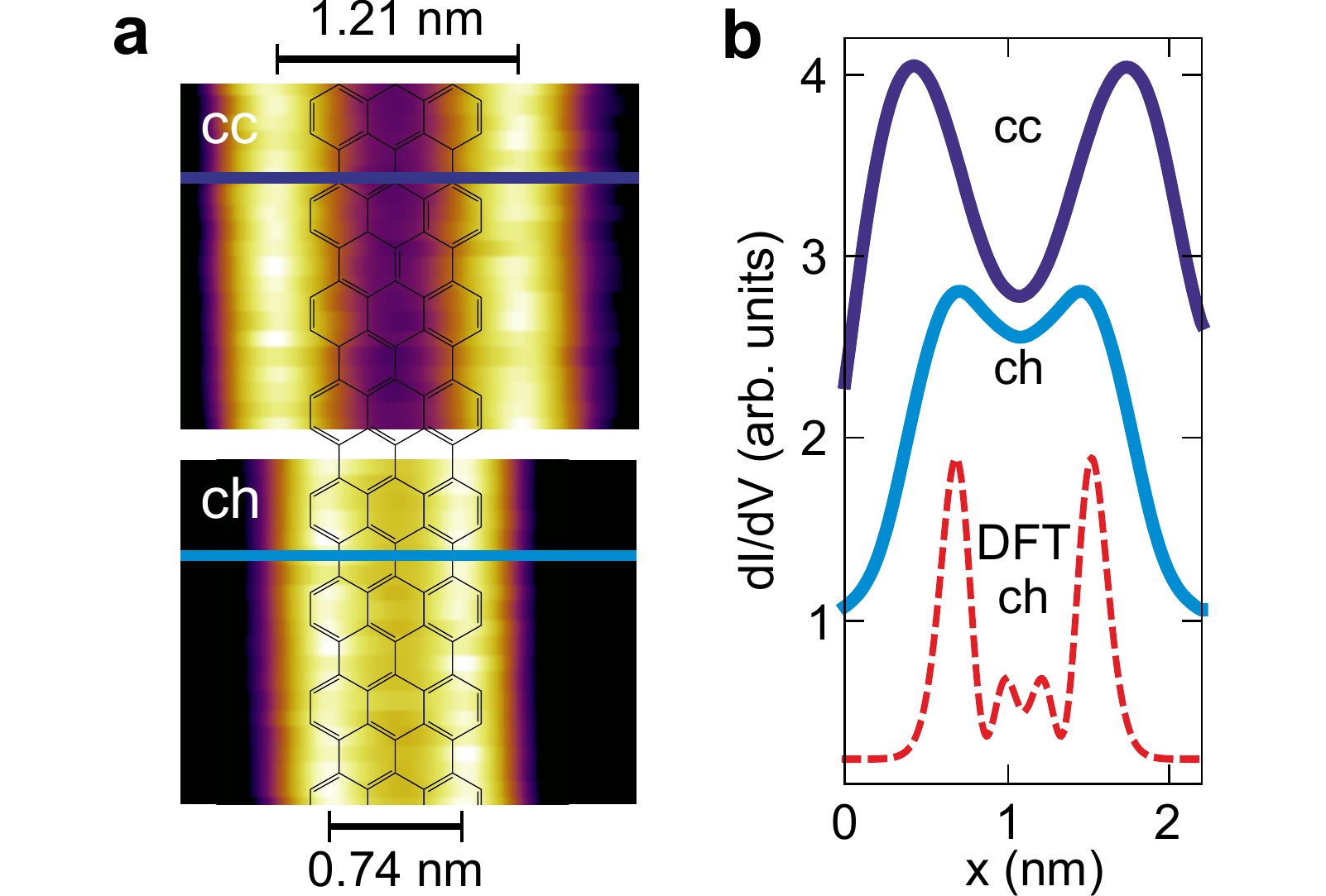}
\caption{\label{fig:4}  (a) Constant-current (cc) and constant-height (ch) d$I$/d$V$ maps revealing different separations between conductance maxima  ($V=2.5$~V, $I=0.3$~nA).
(b) Line profiles as indicated in (a) plus line profile of constant-height DFT-based d$I$/d$V$ simulation at $0.3\,nm$ tip-sample distance ($U=1V$ above the CB onset).}
\end{figure}

Figure~\ref{fig:3} (b) also reproduces the concentration of the LDOS at the edges of the 7-AGNR that has been reported in previous experimental works \cite{Ruffieux2012,Koch2012} and is shown in Fig.~\ref{fig:4} (a). The effect is explained straightforwardly by a lack of cancellation of positive and negative regions of the wave function at the edge of the 7-AGNR.
We note that the large spatial separation of $1.2$~nm observed between the maxima is due to the specific height trajectory followed by the STM tip in constant-current mode. As illustrated in Fig.~\ref{fig:4}, the observed separation reduces to $0.7$~nm in constant-height mode, in agreement with simulations.

\subsection{Band gap extraction}

\begin{figure}[h]
\includegraphics[width=0.7\textwidth]{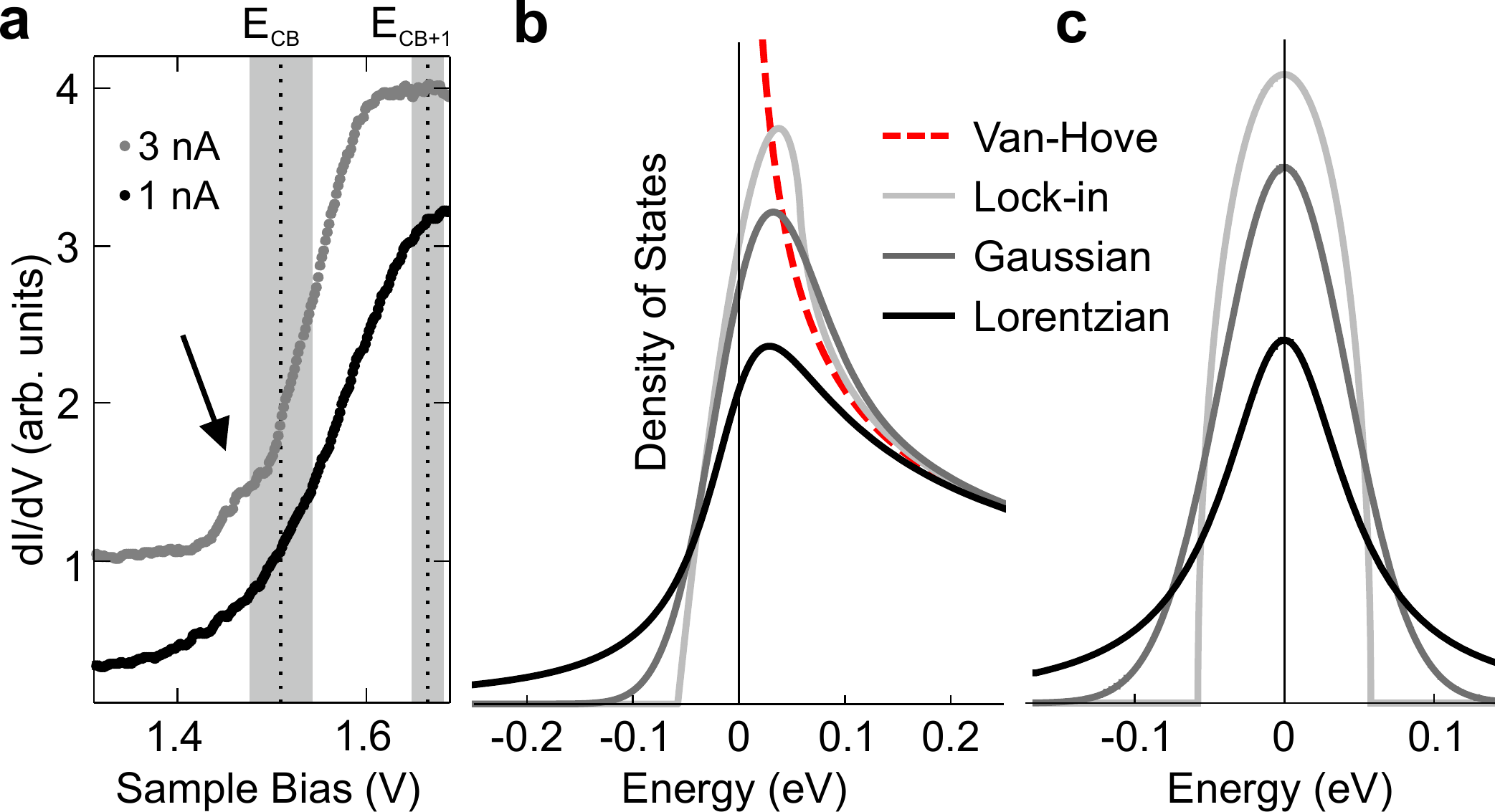}
\caption{\label{fig:5} (a) d$I$/d$V$ spectra at tip-sample distances stabilized on the long edge of the 7-AGNR with $I=1.0$~nA (black) and $3.0$~nA (gray), the latter showing an additional contribution near $E_{CB}$ (see arrow) due to a reduction of the tip-sample distance by $0.13$~nm. The dotted lines with error margins indicate $E_{CB}$ and $E_{CB+1}$ determined from FT-STS.
(b) $1/\sqrt{E}$ Van-Hove singularity (red) with Lock-in, Gaussian and Lorentzian broadening. The broadening functions shown in (c) are normalized to the same value and have $100\,\textrm{meV}$ full-width at half-maximum.}
\end{figure}

The weak signal related to the CB makes it challenging to accurately determine the band gap in STS experiments without momentum resolution.
As shown in Fig.~\ref{fig:5}(a), upon decreasing the tip-sample distance experimentally, a faint additional peak appears in the d$I$/d$V$ spectrum in the range between $1.43$~V and $1.50$~V, which is compatible with a CB onset at $1.53$~eV.
We point out that in Ref. \citenum{Ruffieux2012} and Ref. \citenum{Chen2013} spectra were taken at a lower set-point current and thus at larger tip-sample distance. In these spectra the peak is not detected, leading to a higher apparent band gap, which corresponds to the separation of VB and CB+1.

But even when the signals are sufficiently strong, fitting band onsets from a single spectrum is not straightforward. 
The DOS of a one-dimensional system diverges like $1/\sqrt{E}$ for a band onset at $E=0$. 
Fig.~\ref{fig:5} (b) shows such a van-Hove singularity and its convolution with common types of broadening functions (mathematical details in Appendix B). Lock-in broadening arises from the finite modulation amplitude of the lock-in amplifier used to detect the derivative $dI/dV$. With a peak-peak modulation of $\approx 56\,\textrm{meV}$, this is not the primary source of broadening in this work.
Coupling between the adsorbate and the metal substrate introduces a life-time broadening, which can be modeled by a convolution with a Lorentzian function of appropriate width. Contrary to the case of carbon nanotubes, which exhibit a weaker overall coupling with the substrate \cite{Lin2010}, this broadening can be of the order of $100$~meV for GNRs and different recipes have been used in the literature to extract band gaps: While Chen et al.\ \cite{Chen2013} report the peak-to-peak distance ($E_g = 2.5$~eV), Ruffieux et al.\ \cite{Ruffieux2012} report the distance between half maxima ($E_g = 2.3$~eV).

As illustrated by Fig.~\ref{fig:5} (b), for purely Lorentzian broadening, the actual band onset lies in between half-maximum and maximum (see Appendix B for details). But the exact shape of a spectrum is influenced by other factors, such as the energy-dependence of the spatial exension of the electronic orbitals as well as the energy-dependence of the life-times \cite{Braun2002}.
FT-STS circumvents the problem of having to extract band onsets from a spectrum of complicated shape: the momentum-resolved DOS at finite wave vectors shows peaks with well-defined maxima, allowing for a straightforward extrapolation towards $k=0$.
Even in the case of the weak signal from the CB, which may be missed completely in standard STS analysis, the extrapolation can still be performed.
Using this method, we find $E_g = 2.37 \pm 0.06$~eV for the 7-AGNR on Au(111).

\section{Conclusions}
Despite the high quality of bottom-up fabricated GNRs, the characterization of their electronic properties remains challenging, as exemplified by the wide spread of reported electronic band gaps and effective band masses. For the case of 7-AGNRs, we demonstrate that FT-STS is a robust way of determining the energy dispersion of frontier electronic bands and find effective masses $0.41 \pm 0.08\,m_e$ for the VB, $0.40 \pm 0.18\,m_e$ for the CB and $0.20 \pm 0.03\,m_e$ for the CB+1. While previous reports have been based on single ex-situ measurements, our experiments were performed in-situ with consistent results in several measurements. Moreover, the combination with ab initio electronic structure calculations allows for the unambiguous assignment of the bands and for a detailed understanding of their wave-vector- and height-dependent contribution to the tunneling current. Based on this analysis we show that hitherto, the lowest unoccupied band (CB) has not been detected experimentally and find a band gap of $2.37 \pm 0.06$~eV for metal-absorbed 7-AGNRs. We expect our combined approach to be suitable for the electronic characterization of many kinds of atomically precise armchair and zigzag GNRs with widths in the low nm range.

\section{Acknowledgments}
This work was supported by the Swiss National Science Foundation, the State Secretariat for Education, Research and Innovation via the COST Action MP0901 "NanoTP", the Swiss National Supercomputing Centre (CSCS) under project ID s507, and by the Office of Naval Research BRC Program.

\bibliography{library}

\appendix
\section{FT-STS on 7-AGNRs of different lengths}

\begin{figure}[h]
\includegraphics[width=0.9\textwidth]{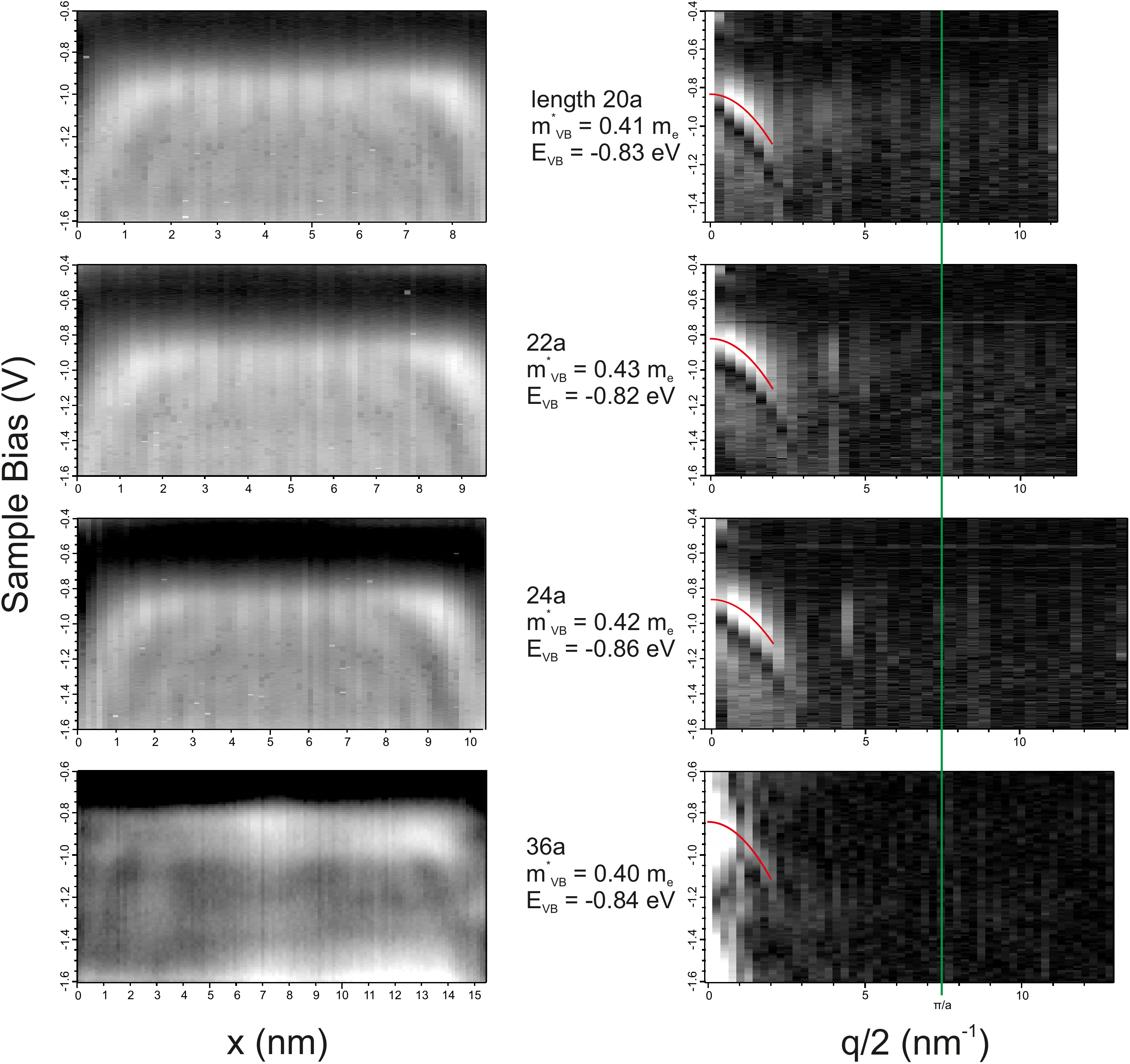}
\caption{\label{fig:FT-STS} Equidistant d$I$/d$V(V)$ spectra (left) and corresponding line-by-line Fourier transform (right) of occupied states, recorded along the edge of several 7-AGNRs.
Indicated are the length of the GNRs, together with the values for $m^*$ and $E(k=0)$ extracted from the parabolic fits (red lines). The green lines indicate the border of the first Brillouin zone at $q/2=\frac{\pi}{a}\approx 7.3\,nm^{-1}$.}
\end{figure}

The values for $E(k=0)$ and $m^*$ in Table \ref{tab:1} in the main text are averaged over several measurements on 7-AGNRs of different lengths. The corresponding values are listed in Table~\ref{tab:me}, including the ones shown in Fig.~\ref{fig:FT-STS} and Fig.~\ref{fig:2} in the main text. Despite possible influences of systematic errors, specifically the tip shape, the setpoint of the spectra and the length of the GNRs, the obtained results for $E(k=0)$ and $m^*$ are consistent. The parabolic fit for the determination of $m^*$ and $E(k=0)$ was performed in the k-range  [$0, 2\ nm^{-1}$] for the VB and CB, and [$0, 1.5\ nm^{-1}$] for the CB+1. 

As pointed out in the main text, the dominant source of uncertainty is the selection of the real-space window for the discrete Fourier transform of the scanning tunneling spectra. For each set of spectra, the width of the window has been varied and the relative deviations in the effective masses, ranging from 15\% to 43\%, have been noted. Once the respective intensity maxima at each k-value are determined, the error in the parameters of the parabolic fit are below 2\% for the onset energy and below 5\% for the effective masses. Due to the predominance of the window selection, the errors of the fit parameters have been neclegted for the averaged values. The uncertainties given in Table~\ref{tab:me} result from the propagation of the uncertainties in the individual spectra to the averaged values.

\begin{table}[h]
  \centering
\begin{tabular*}{0.75\textwidth}{@{\extracolsep{\fill}} c | c  c | c  c | c  c @{\extracolsep{\fill}}}
  Length & $m^*_{VB}$ & $E_{VB}$ & $m^*_{CB}$ & $E_{CB}$ & $m^*_{CB+1}$ & $E_{CB+1}$ \\ \hline
  20$a$ & $0.41\ m_e$ & $-0.84\ eV$ &  &  &  &  \\
  20$a$ & $0.41\ m_e$ & $-0.83\ eV$ &  &  &  &  \\
  22$a$ & $0.43\ m_e$ & $-0.82\ eV$ &  &  &  &  \\
  22$a$ & & & $0.41\ m_e$ & $1.52\ eV$ &  $0.21\ m_e$ & $1.66\ eV$  \\
  22$a$ & & & & &  $0.20\ m_e$ & $1.67\ eV$  \\
  24$a$ & $0.42\ m_e$ & $-0.86\ eV$ &  &  &  &  \\
  24$a$ & & & $0.40\ m_e$ & $1.52\ eV$ &  $0.20\ m_e$ & $1.67\ eV$  \\
  26$a$ & & & $0.40\ m_e$ & $1.51\ eV$ &  $0.20\ m_e$ & $1.67\ eV$  \\
  36$a$ & $0.40\ m_e$ & $-0.84\ eV$ &  &  &  &  \\
\end{tabular*}
  \caption{\label{tab:me}Effective masses $m^*$ and band onsets $E(k=0)$ of all measurements included in the averages reported in Table~\ref{tab:1}. The length is given in units of $a=0.43\ nm$, the unit cell length}
  \label{tab:myfirsttable}
\end{table}

\clearpage

\section{Broadening of the density of states}

In one dimensional periodic systems, such as graphene nanoribbons,
band onsets are characterized by a van-Hove singularity
in the density of states.
In the following, we consider the convolution of a function

\[ f(E) = \left\{\begin{array}{ll} \frac{1}{\sqrt{E}} & E > 0 \\
                            0                  & E \leq 0
          \end{array}\right.\quad, \]

representing the density of states for a band onset at $E=0$, with a broadening function $g(E)$ to obtain the broadened density of states
\[  h(E)  = \int_{-\infty}^\infty f(E') g(E-E')\,dE'\quad. \]

\begin{figure}[h]
    \centering
    \includegraphics[width=0.7\textwidth]{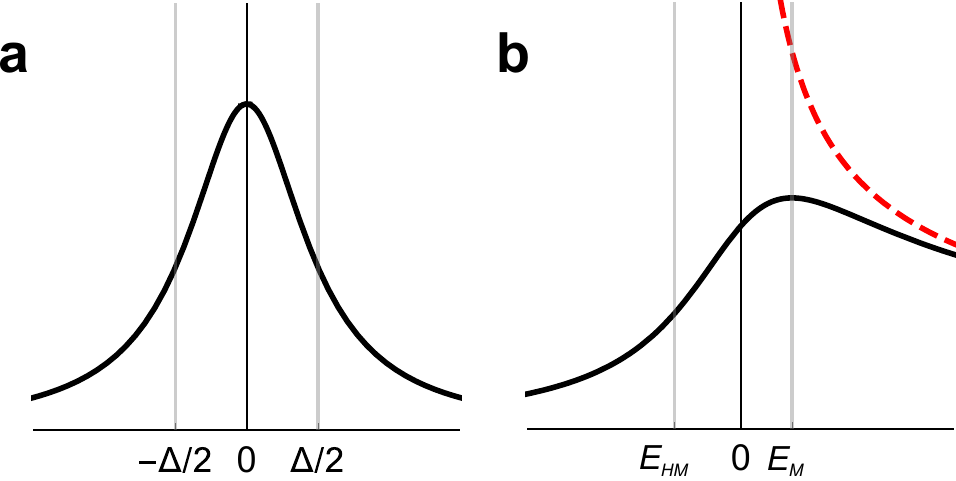}
    \caption{(a) Lorentzian function with $\Delta$ full-width at half-maximum. (b) Convolution $h(E)$ (black) of van-Hove singularity $f(E)$ (dashed red) with Lorentzian. The positions of half-maximum ($E_{HM}$) and maximum ($E_M$) are indicated.}
    \label{fig:comp}
\end{figure}

Figure \ref{fig:comp} illustrates the convolution of $f(E)$ with a Lorentzian function of $\Delta$ full-width at half-maximum (FWHM). 
The broadened density of states $h(E)$ assumes its half-maximum at $E=E_{HM}$ \emph{below} the band onset, while the maximum is assumed at $E=E_M$ \emph{above} the band onset. This also holds for the other types of broadening considered here (see Figure \ref{fig:5}(b) in main text). Using either $E_{HM}$ or $E_M$ to determine the band onset therefore introduces a systematic bias in the resulting band gaps.

If $g(E,\Delta)$ denotes a broadening function with FWHM $\Delta$, then $g(E,1) \propto g(E\Delta, \Delta)$ and also $h(E,1) \propto h(E\Delta, \Delta)$.
The energies $E_{HM},E_M$ are thus proportional to the FWHM  of the broadening function and table \ref{tab:energies} presents the corresponding proportionality factors.
Mathematical details are provided in the following.


\begin{table}
    \centering
    \begin{tabular}{cccc}
        & Lorentzian & Lock-in & Gaussian \\\hline
        $E_M$ [$\Delta$] & 0.289 & 0.377 & 0.325 \\
        $E_{HM}$ [$\Delta$] & -0.371 & -0.258 & -0.296
    \end{tabular}
    \caption{Numerical values of positions $E_M$ of the maximum and $E_{HM}$ of the half-maximum of $h(E)$, given in units of the full-width at half-maximum $\Delta$ of the broadening function. Band onset is at $E=0$.}
    \label{tab:energies}
\end{table}


\subsection{Lorentzian broadening} \label{sec:lorentz}

The coupling between molecules and the underlying substrate gives rise
to a finite life time of excited states, leading to a Lorentzian broadening of the corresponding energy levels.

For a Lorentzian broadening
\[ g(E) = \frac{1}{\pi} \frac{\Gamma}{2} / \left( E^2 + (\Gamma/2)^2 \right) \]
with FWHM $\Delta = \Gamma$, we obtain
\begin{eqnarray*}
    h(E) &=&
    \frac{1}{\sqrt{2}} \left(\frac{1}{\sqrt{2E+i\Gamma}} + \frac{1}{\sqrt{2E-i\Gamma}} \right) \\
    &=& \frac{\sqrt{ 1 + \textrm{sgn}(E) / \sqrt{1+ (\Gamma/2E)^2} }}{
    \sqrt[4]{ 4E^2+\Gamma^2}}
\end{eqnarray*}
which assumes its maximum at $E_M= \Gamma / (2\sqrt{3}) \approx 0.289\, \Delta$.

Half-maximum is assumed at
\begin{eqnarray*}
    E_{HM} &=& - \frac{1}{3} \sqrt{\frac{229}{12} + \frac{4}{3} \beta
         - \frac{4}{3} \sqrt{494 - 9/\alpha - 9\alpha + 7760/\beta }}\ \Gamma\\
         \textrm{where}\quad  \alpha &=& \sqrt[3]{31 + 8 \sqrt{15}}\quad,\\
         \beta &=& \sqrt{247 + 9/\alpha + 9 \alpha} \quad.
\end{eqnarray*}
with numerical value $E_{HM} \approx -0.371\, \Delta$.

\subsection{Lock-in broadening}\label{sec:lock-in}

Experimentally, the derivative of the tunneling current $I$ with
respect to the bias voltage $V$ is approximated by the lock-in derivative

\begin{eqnarray*}
    \frac{dI}{dV}(V,\delta V) &\propto&
        \int_0^{2\pi/\omega} \cos(\omega t)\, I\left(V+ \frac{\delta V}{2} \cos(\omega t)\right) dt
\end{eqnarray*}

The bias voltage $V$ is modulated with a reference signal $\frac{\delta V}{2} \cos(\omega t)$ and  the time-integral of the product between tunneling current $I(V)$ and reference signal is recorded. In the limit $\delta V\rightarrow 0$, the exact derivative $dI/dV(V)$ is recovered.

Setting $\omega=1$ for convenience, the expression for $\frac{dI}{dV}(V,\delta V)$ can be transformed into a convolution of the exact derivative $\frac{dI}{dV}(V)$ with a broadening function:

\begin{eqnarray*}
    \frac{dI}{dV}(V,\delta V)
    &\propto & \int_0^{2\pi} \cos(t)\, I\left(V+ \frac{\delta V}{2} \cos(t)\right) dt \\
    &\stackrel{x=\cos t}{=}&
     2 \int_{-1}^1 \frac{x}{\sqrt{1-x^2}}\, I\left(V+\frac{\delta V}{2} x\right) dx\\
    &=&
        \left.2 I\left(V + \frac{\delta V}{2}x \right) \left(- \sqrt{1-x^2} \right)\right|_{x=-1}^1 \\
    &&
       -2 \int_{-1}^1 \left(- \sqrt{1-x^2} \right) \frac{\delta V}{2} \frac{dI}{dV}\left(V+\frac{\delta V}{2}x\right)dx \\
    &=& \delta V \int_{-1}^1 \sqrt{1-x^2} \frac{dI}{dV}\left(V+\frac{\delta V}{2}x\right)dx  \\
    &\stackrel{y=\frac{\delta V}{2} x}{=}&
    2 \int_{-\delta V/2}^{\delta V/2} \sqrt{1-\left(\frac{2y}{\delta V}\right)^2}
                   \frac{dI}{dV}\left(V+y\right)dy
\end{eqnarray*}

After normalization, we obtain
\[ g(E) = \left\{ \begin{array}{cc} 
      \frac{4}{\pi} \frac{1}{e \delta V} \sqrt{1-\left(\frac{2E}{e \delta V}\right)^2} & -\frac{e\delta V}{2} < E < \frac{e\delta V}{2} \\
       0 & otherwise \end{array}\right.\]
with FWHM $\Delta = \frac{\sqrt{3}}{2}e \delta V$.

In lack of an analytical solution for the convolution, we present numerical values. $h(E)$ assumes its maximum at $E_M \approx 0.326115\, e\delta V \approx 0.377\, \Delta$, and the half-maximum at $E_{HM} \approx -0.223072\, e \delta V \approx -0.258\, \Delta$.

\emph{Note:} Lock-in broadening has not been considered in the STS simulations, since the experimental peak-peak modulation of $\delta V=2\cdot 20\,\textrm{mV}\cdot \sqrt{2}\approx 56\,\textrm{mV}$ was significantly smaller than the effective broadening observed. It was verified that the effect of lock-in broadening on the simulated spectrum can be neglected.

\subsection{Gaussian broadening}\label{sec:gauss}

For completeness, we also provide results for Gaussian broadening
\[ g(E) = \frac{1}{\sqrt{2\pi}\sigma} \exp\left( - \frac{E^2}{2\sigma^2} \right) \]
with FWHM $\Delta = \sqrt{8\ln(2)}\, \sigma$.
For the convolution, we obtain

\begin{eqnarray*}
    h(E) &=& \frac{1}{\sqrt{2\pi} \sigma} \frac{\pi}{2} \sqrt{|E|}
              \exp\left( - \frac{E^2}{4\sigma^2} \right)  \left[
                   I_{-\frac{1}{4}}\left( \frac{E^2}{4\sigma^2} \right)
               + \textrm{sgn}(E)  I_{\frac{1}{4}}\left( \frac{E^2}{4\sigma^2} \right) \right]
\end{eqnarray*}

where $I_\alpha(x)$ denotes the modified Bessel function of the first kind.

In lack of analytical expressions, for $E_M$ and $E_{HM}$, we present numerical values. The maximum of $h(E)$ is assumed at $E_M \approx 0.764951\, \sigma \approx 0.325\, \Delta$, the half-maximum at $E_{HM} \approx - 0.697669\, \sigma \approx -0.296\,\Delta$.

\end{document}